\documentclass[aps, prb, reprint]{revtex4-1}

\usepackage{graphicx}

\newcommand{\BS}{Bi$_{2}$Se$_{3}$}
\newcommand{\bvec}[1]{\boldsymbol{\mathrm{#1}}}
\usepackage{amsmath}

\begin{document}

\title{Absence of Magnetic Fluctuations in the Ferromagnetic/Topological Heterostructure EuS/Bi$_{2}$Se$_{3}$}

\author{Gavin Osterhoudt}
\author{Ryan Carelli}
\author{Kenneth S. Burch}
\email{ks.burch@bc.edu}
\affiliation{Department of Physics, Boston College, 140 Commonwealth Ave Chestnut Hill, MA 02467-3804, USA}

\author{Ferhat Katmis}
\author{Nuh Gedik}
\author{Jagadeesh Moodera}
\affiliation{Department of Physics, MIT, 77 Massachusetts Avenue, Cambridge, MA 02139-4307, USA}

\date{\today}

\begin{abstract}
Heterostructures of topological insulators and ferromagnets offer new opportunities in spintronics and a route to novel anomalous Hall states. In one such structure, EuS/\BS\, a dramatic enhancement of the Curie temperature was recently observed. We performed Raman spectroscopy on a similar set of thin films to investigate the magnetic and lattice excitations. Interfacial strain was monitored through its effects on the \BS\ phonon modes while the magnetic system was probed through the EuS Raman mode. Despite its appearance in bare EuS, the heterostructures lack the corresponding EuS Raman signal. Through numerical calculations we rule out the possibility of Fabry-Perot interference suppressing the mode. We attribute the absence of a magnetic signal in EuS to a large charge transfer with the \BS. This could provide an additional pathway for manipulating the magnetic, optical, or electronic response of topological heterostructures.
\end{abstract}


\maketitle

Since the prediction\cite{zhang2009} and verification\cite{chen2009Science,hasan2009PRL} of topological surface states in \BS\ there has been significant interest in the creation of heterostructures involving these topological insulators (TI). One promising combination is found in the joining of thin film ferromagnets with TI's. The exchange field present at the interface of such a heterostructure breaks the time reversal symmetry in the TI surface states\cite{gedik2016NCom}, opening a gap\cite{yang2013PRB,kandala2013APL,li2015PRL,katmis2016,chen2010Sci}. Such a system has been seen to display negative magnetoresistance\cite{yang2013PRB} and the quantum anomalous Hall effect\cite{chang2013Sci,kou2015SSC}, Furthermore, the strong spin-momentum locking leads to a large spin-torque which may prove useful for spintronics applications\cite{ralph2014Nat,li2014NNano}.

Other progress in the development of applications with heterostructures has been enabled by the charge transfer phenomenon. A rewritable nanoscale metal to insulator transition making use of this has been demonstrated in LAO/STO\cite{levy2008NatMater}. Beyond enabling new functionalities, charge transfer can also enhance preexisting effects. In FeSe a large charge transfer has been shown to enhance the superconducting transition temperature by nearly an order of magnitude\cite{wang2016SST}. To date there has been no evidence presented of charge transfer enabling new effects in topological insulators. In this letter we present evidence for a large charge transfer in a ferromagnetic/topological heterostructure with dramatically enhanced Curie temperature (T$_{C}$).

A recent report by some of us presented evidence that in a heterostructure of \BS\ and the ferromagnetic insulator (FI) EuS, the T$_{C}$ of EuS was increased by over an order of magnitude, leading to traces of magnetization still present at room temperature\cite{katmis2016}. Despite the exciting possibility for room temperature devices and novel physical effects, the interface and resulting changes in both materials are still poorly understood. Two important aspects that need to be addressed are the changes in the lattice and magnetic excitations due to the interface. The lattice could affect the resulting magnetism through the inverse magnetostriction effect\cite{hellman2016} and the phonons are an important factor in the transport properties of TI's\cite{reijndersPRB2014,wang2012PRL,kim2012PRL}. Likewise, understanding the magnetic excitations can provide insight into the dynamics at the interface. In order to address both of these simultaneously we have used Raman spectroscopy, which has successfully tracked magnetic and lattice excitations in other heterostructures and 2D materials\cite{sandilands2010PRB,ferrari2013NNano,wang20162DM,lee2016NanoL,tian20162DM}. For the case examined here, it is well established that the Raman spectra of EuS are sensitive to the presence of magnetic ordering\cite{schlegel1973SSC}. Through one spectral measurement we should therefore be able to probe both the phonon structure and the magnetic ordering in nano-scale \BS/EuS heterostructures.

\begin{figure}
\begin{center}
\includegraphics{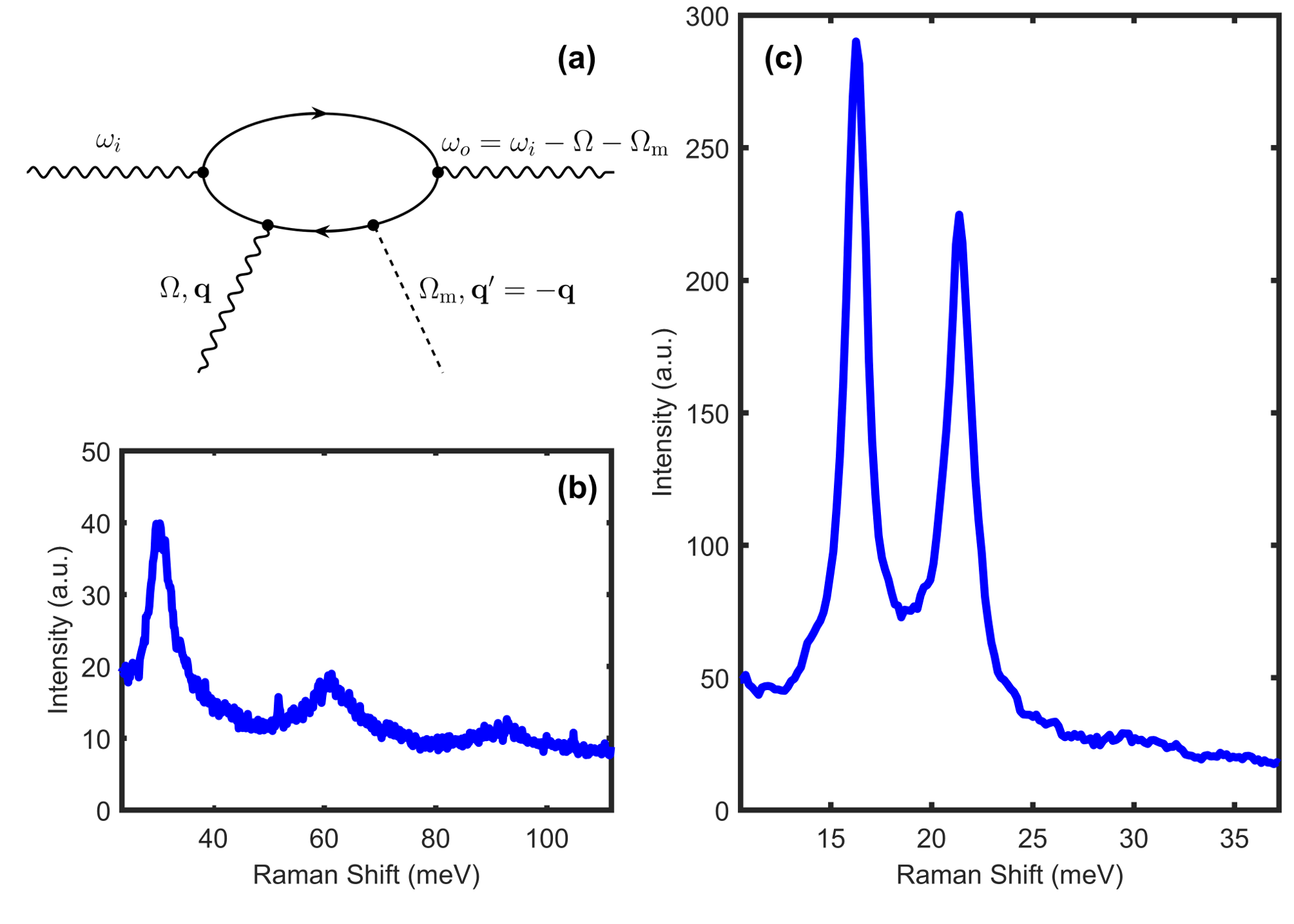}
\end{center}
\caption{(a) The Feynman diagram for the Raman scattering process in EuS. Incident light first generates an electron--hole pair. The hole scatters with both the lattice and spin system before recombination. (b) Room temperature Raman spectra from 5 nm thick EuS on sapphire. The fundamental mode is observed at 30.4 meV, with the second harmonic at 60.8 meV. The third harmonic is weakly visible at 91.2 meV. (c) Raman spectra of 7QL thick \BS\ with 10 nm EuS grown on a sapphire substrate. The EuS mode at 30.4 meV is absent in measured heterostructures.}\label{Figure1}
\end{figure}

At first glance one should expect little from the Raman spectra of EuS. The crystal structure of EuS has two inter--penetrating face--centered cubic lattices and is symmetric under inversion. However, there is no \emph{unique} center of inversion, and thus no modes that are even under inversion. A group theoretical analysis of the phonons in such a lattice reveals that at the zone center the optical modes are of $ T_{1u} $ symmetry and therefore Raman inactive. Despite this, in the EuX family of compounds (X = O, S, Se, Te) first--order Raman scattering is experimentally observed\cite{merlin1977PRL,schlegel1973SSC,grunberg1977PB,guntherodt1979PRB} as seen in FIG.~\ref{Figure1}(b). The underlying mechanism behind this symmetry forbidden scattering has its origins in the spin--disorder present in the paramagnetic phase\cite{merlin1977PRL}. While phonons from throughout the Brillouin zone are not typically excited by optical means due to conservation of momentum, in these materials the disordered spin system is capable of providing the necessary momentum to balance out the phonon contribution\cite{guntherodt1979PRB}. In particular, the LO phonon mode at the L point in the Brillouin zone has been shown to be excited in this scattering event\cite{grunberg1977PB,guntherodt1979PRB}. The Feynman diagram for such a process is shown in FIG.~\ref{Figure1}(a). The incoming photon of energy $ \hbar \omega_{i} $ excites an electron from the localized 4f valence band to the 5d conduction band, leaving behind a hole. The hole then interacts with both the lattice and the spin system, producing an LO phonon of energy $ \hbar \Omega $ and momentum $ \bvec{q} $, and a spin excitation with energy $ \hbar \Omega_{m} $ and momentum $ \bvec{q}' $. The recombination of the electron and hole pair then emits the Raman scattered light of energy $ \hbar \omega_{o} $. Note that, although the spin system provides momentum $ \bvec{q}' = - \bvec{q} $, there is no energy cost associated with changing the spin since different spin states are degenerate in the paramagnetic phase (i.e., $ \hbar \Omega_{m} = 0 $). When the crystal has long--range magnetic order, the magnons at finite $\bvec{q}$ now require finite $ \hbar \Omega_{m} $. However, the probability of creating a magnon via the Raman process is proportional to $ ( \hbar \Omega_{m})^{-1} $. Therefore as the system approaches and passes through its Curie temperature, the intensity of the Raman scattering is quenched\cite{schlegel1973SSC,guntherodt1979PRB}.

Since this is a higher order scattering process involving the excitation of both a phonon and a spin, the mode intensity might be expected to be weak. However, such a process is also highly resonant with the excitation energy. By tuning the laser to the right intermediate electronic state, the intensity of the EuS mode is strongly enhanced. Indeed, measurements of the scattering intensity as a function of excitation wavelength indicate that there is a strong resonance effect which has a maximum in the $\sim$2.2 eV range\cite{merlin1977PRL}. Alternatively, we can achieve a similar enhancement (or suppression) by tuning the Fermi energy and changing the states involved in the optical transition. As a result of this, either through entering the magnetically ordered state,  tuning the intermediate state by choice of laser \emph{or} doping we can eliminate the Raman scattering from EuS.

The reported magnetic order at room temperature in EuS/\BS\ heterostructures should thus also lead to a strong suppression of the magnetic Raman signatures at ambient temperatures. We explored this possibility with a series of EuS/\BS\ samples with varying EuS thicknesses grown on two different substrates. This allowed us to also investigate the role of strain in these samples. Raman spectra were acquired using a WITec alpha300R confocal Raman system (detailed information about the synthesis of the heterostructures may be found in\cite{katmis2016}). A 100X objective was used to focus the unpolarized, 532 nm (2.33 eV) light down to a ~1 $\mathrm{\mu}$m spot size. A power of 10 $\mathrm{\mu}$W was used to avoid local heating of the \BS\cite{burch2011APL}. Unphysical artifacts from ``cosmic rays" were removed using an algorithm based on wavelet transformations and data clustering methods\cite{tian2016AS}. Spectra that have had cosmic rays removed were averaged and normalized by power and integration time.

At room temperature, bare EuS should be in the paramagnetic phase and display a measurable Raman response. The results of our room temperature measurements on a film of 5 nm thick EuS (sapphire substrate) are shown in FIG.~\ref{Figure1}(b). As previously observed in the bulk, the spectrum clearly displays the fundamental mode at 30.4 meV as well as the second and third harmonic overtones (at 60.8 and 91.2 meV respectively). Our measured value for the fundamental is slightly higher than the typical room temperature value of 29.8 meV\cite{schlegel1973SSC,merlin1977PRL,grunberg1977PB}. While the reason for this shift is unclear, the ease of observation of the mode is consistent with the expectation of paramagnetism at ambient temperatures.

Surprisingly we find that the \BS/EuS heterostructures lack the EuS mode. In FIG.~\ref{Figure1}(c) we show the spectra acquired from a sample with 10 nm thick EuS and 7 QL thick \BS. While there are two peaks which belong to the \BS\ (these are discussed later) the EuS mode is conspicuously absent. The thickness of the EuS layer is double that of the bare EuS sample, yet we still do not observe any corresponding Raman signatures. This is the case for every heterostructure that we measured, regardless of EuS or \BS\ thickness. One possible explanation is the previously reported room temperature ferromagnetism has suppressed the magnetic excitation. Alternatively, there are two physical effects unrelated to the magnetism that may alter the Raman spectra.

The Raman from EuS could be suppressed by modifications of the electronic system. In the typical EuS Raman scattering process an electron is excited to the 5$d$ conduction band and leaves a hole behind in the localized 4$f$ valence band. The rest of the scattering process then proceeds as illustrated in FIG.~\ref{Figure1}(a). This process is highly resonant upon laser excitation energy with an amplification of two orders of magnitude occurring at our excitation energy of 2.33 eV, which is very nearly at the maximum of the resonance\cite{merlin1977PRL}. Such resonance is typically observed when the transitions in the scattering process involve \emph{real} energy levels instead of virtual energy levels\cite{cardona1983}. An implicit assumption in this discussion has been that the EuS is undoped and the Fermi level lies within the band gap of the EuS. However, if the EuS were to become doped and the Fermi level shifted into either the conduction or valence bands then we would have a change in the energy levels involved in the scattering process.

\begin{figure}
\begin{center}
\includegraphics{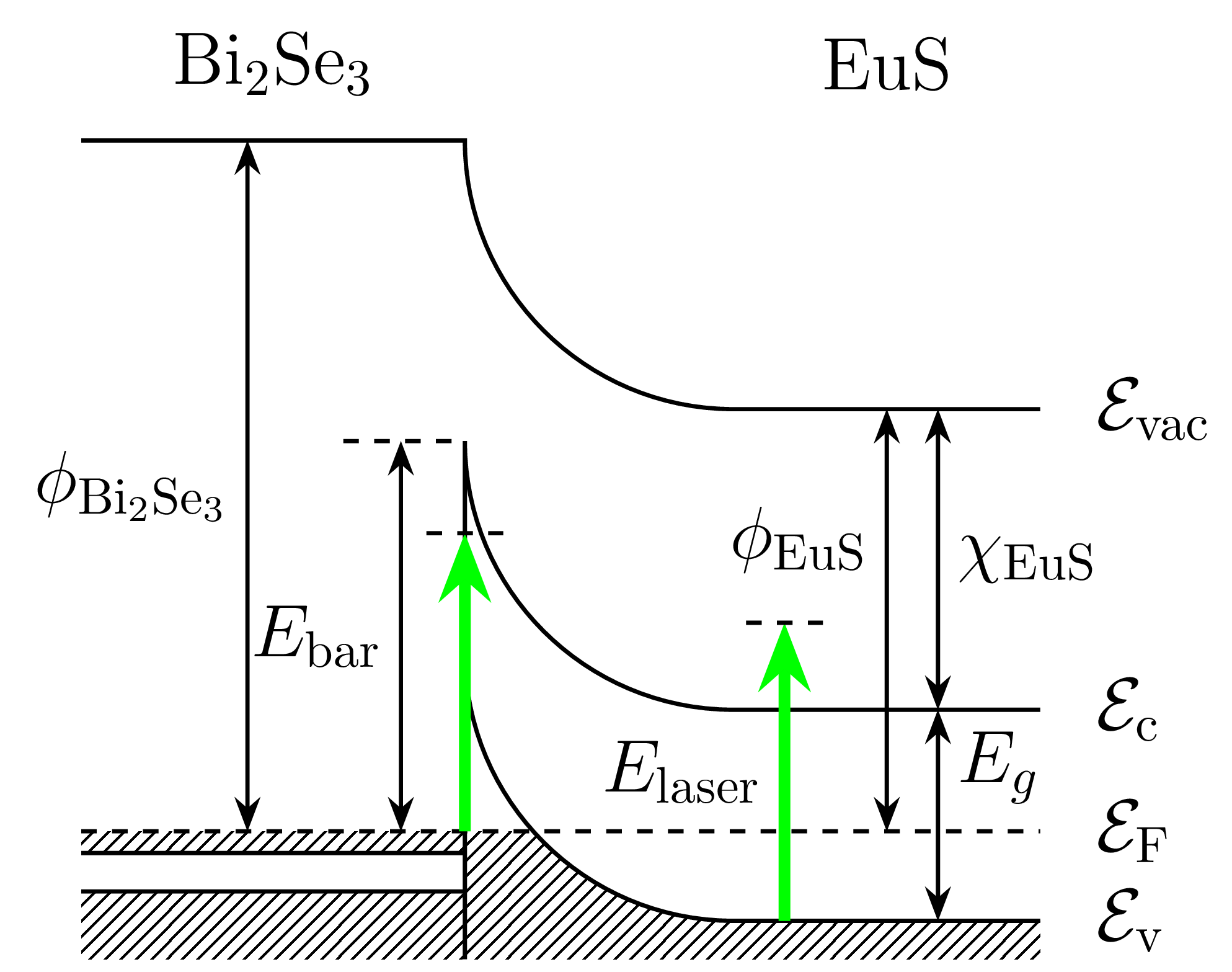}
\end{center}
\caption{Schematic diagram of the band bending that occurs at the interface between the metallic \BS\ and the semiconducting EuS. The work function of \BS\ ($\phi_{\text{\BS}}$ = 5.4 eV) is larger than that of EuS ($\phi_{\text{EuS}}$ = 3.3 eV) which leads to the formation of a Schottky barrier of height $ \phi_{\text{\BS}} - \chi_{\text{EuS}} = 3.05 $ eV at the interface. To balance out the chemical potential electrons move from the EuS into the \BS\ which lowers the Fermi level within the depletion region into the valence band of EuS. While photoexcitation of the EuS with a 2.33 eV laser is possible in the bulk material (band gap $E_{g} = 1.65 $ eV), in the depleted region the difference between the Fermi energy and the conduction band minimum is larger than 2.33 eV.}\label{Band_Bending}
\end{figure}

In fact, such a change in Fermi level is possible due to the band bending and the charge transfer that occurs at the \BS/EuS interface. In a simple model we may treat their interface as a metal--semiconductor junction, since the \BS\ is sufficiently $n$-doped ($n\sim10^{19}$ cm$^{-3}$) placing the Fermi energy deep in the conduction band\cite{katmis2016}. The amount that the EuS bands will bend at the interface depends on the difference in work function between the two materials. \BS\ has a work function of 5.4 eV and EuS has a work function of 3.3 eV (and electron affinity of 2.35 eV)\cite{edmonds2014ACSN,eastman1969PRL}.

In FIG.~\ref{Band_Bending} we show a schematic of the band bending that occurs as a result of this mismatch. In order for the chemical potential at the interface between the two materials to be equal, electrons flow out of the EuS into the \BS. The barrier height formed at the interface is found as the difference between the work function of the \BS\ and the EuS: $E_{\text{bar}} = \phi_{\text{\BS}} - \chi_{\text{EuS}} = 3.05 $ eV. The difference between the barrier height and the 1.65 eV band gap in EuS tells us that the built-in potential is 1.4 eV. In other words, the Fermi level in the EuS is shifted down into the valence band by 1.4 eV. The spatial extent of this depleted layer is expected to be on the order of hundreds of nanometers and since our EuS films are only nanometers thick we expect that the entire layer experiences this effect. In the bare EuS our laser energy of 2.33 eV is capable of exciting electrons across the bandgap of 1.65 eV. However in the \BS/EuS heterostructure, the large shift in chemical potential requires $\approx 3$ eV to optically excite from the valence to conduction bands. The only available transitions for Raman scattering will then involve virtual energy levels, drastically reducing the intensity. This provides a natural explanation for the observed absence of the mode in our spectra. It is possible that future Raman experiments using a laser with higher energy per photon could be performed which would enable the observation of the EuS Raman mode.

Interestingly the band bending and associated introduction of holes into the EuS could be partially responsible for the previously reported increase in T$_{C}$. Furthermore, should magnetic order be present at room temperature this could also explain the lack of a signal from the EuS. Unfortunately attempts to heat the sample above T$_{C}$ to observe the re-emergence of the peak were unsuccessful. Thus further studies are warranted, perhaps with ionic liquid gating or alternate capping layers to further elucidate the role of doping in our results.

A second explanation for the mode's absence is Fabry--Perot interference due to the multi--layered nature of the samples. In a multi-layered structure there is the possibility of multiple reflections interfering with each other and either enhancing or suppressing both the incident laser and Raman scattered radiation\cite{nemanich1980PRL,ramsteiner1989AO}. In order to address the role of interference, we performed numerical calculations that take into account multiple reflections at each interface. Similar calculations have been performed for a wide variety of quasi--two dimensional systems and  were successful in explaining the variation in intensity with the thickness of the dielectric or exfoliated layers\cite{geim2007APL,wang2008APL,yoon2009PRB,zhang2015APL_MRM,sandilands2010PRB,burch2011APL}.

\begin{figure}
    \centering
    \includegraphics{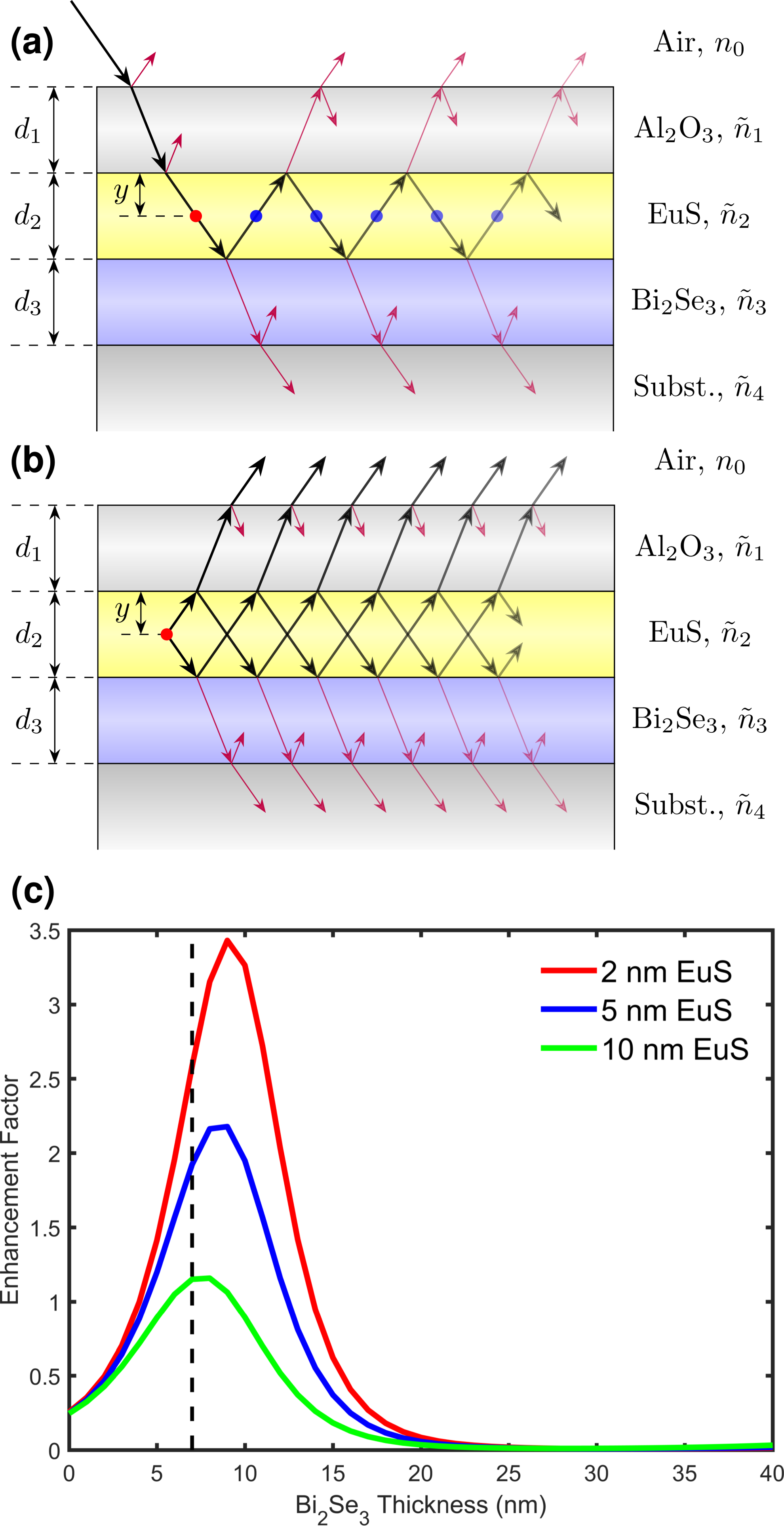}
    \caption{(a) The incident laser undergoes multiple reflections at each interface of the heterostructure. (b) Raman scattered light also undergoes multiple reflections due to the multiple interfaces. The resulting enhancement factor is calculated by integrating the combined effect from each process over the thickness $d_{2}$. The calculation considers light only at normal incidence, the non-normal angles are drawn solely for clarity. (c) The enhancement factor varies with both the thickness of the \BS\ layer and the EuS layer. The thickness of \BS\ in our measured samples is indicated by the dashed line.}\label{Interference}
\end{figure}

We developed an extension of the multi-reflection model (MRM) used by Zhang \emph{et al.}\cite{zhang2015APL_MRM}. A diagram representing each of the two interference processes considered in the calculations is shown in FIG.~\ref{Interference}. The first of these, illustrated in FIG.~\ref{Interference}(a), is the interference of the incident laser with itself while the second, illustrated in FIG.~\ref{Interference}(b), is the interference of the Raman scattered light. The model considers light at \emph{normal} incidence, the rays are drawn at non--normal angles of incidence for clarity only. Details of the derivation for the enhancement factor may be found in the supplemental information. 

In FIG.~\ref{Interference}(c) we show the interference enhancement factor obtained at the energy of the EuS Raman mode as a function of the \BS\ thickness. The Al$_{2}$O$_{3}$ capping layer was fixed at 5 nm thick and the calculation repeated for multiple thicknesses of EuS. In the absence of a \BS\ layer, i.e. when the thickness is zero, we see that we obtain an enhancement factor of 0.25. This indicates that, as expected due to the light exiting the back surface of the sample and the small volume, Raman signal from a thin slab of EuS will be reduced compared with a bulk crystal. In contrast, we see that at the thickness of \BS\ in our samples, indicated by the dashed line, we obtain an enhancement factor of 2.5, 1.9, and 1.2 for EuS thicknesses of 2, 5, and 10 nm respectively. We should therefore expect to see a $2.5/0.25 = 10 $, $1.9/0.25\approx7.6$, or $1.2/0.25\approx4.8$ times larger signal from the AlO$_{x}$/EuS/\BS/sapphire films compared to just AlO$_{x}$/EuS/sapphire. However, as seen in the measured spectra this is clearly not the case. We therefore rule out the possibility of FP interference causing the absence of the EuS Raman mode.

While the charge transfer discussed above seems to provide an explanation for the absence of the EuS Raman mode we can also investigate the role of the lattice in the $ T_{C} $ enhancement through the spectral features associated with \BS. The interfacial strain between EuS and \BS\ produces measurable effects on the phonons of the \BS\ from which we can characterize the nature of the strain as well as the uniformity of the films. The results of our measurements may be found in the supplemental, but to summarize our analysis we find that there are different types of strain experienced by the \BS\ depending on the type of substrate it is grown on. In the case of a sapphire substrate the \BS\ experiences a tensile strain causing the phonon modes to shift down in energy, while with an STO substrate the strain is compressive, causing the phonon modes to shift up in energy. The addition of EuS adds another source of tensile strain that competes with the strain from the substrate, as observed through the dependence of the energy shifts on the thickness of the EuS layer. From Newton's third law we infer that if EuS exerts a tensile strain on \BS\ then the \BS\ must exert a compressive strain on EuS. However, films with various \BS\ thicknesses and substrates displayed similar T$_{C}$ enhancement, despite the large differences in the strain that we observed. This is somewhat surprising, given that a reduction in the lattice constant of EuS is known to increase the strength of the magnetic interactions and lead to higher $ T_{C} $\cite{katmis2016,goncharenko1998PRL,sollinger2010PRB}.

In summary, we have performed a series of Raman measurements on \BS/EuS heterostructures. The EuS Raman mode, which is an indicator of the degree of magnetic ordering, is not observed in heterostructures. While this could be a confirmation of the presence of ferromagnetism in the films at room temperature, there are other effects that could also be obscuring the observation of this mode. Our numerical calculations of optical interference indicate that this is not likely to be the cause, but rather charge transfer from the EuS is suppressing this mode. Such large charge transfer could open an additional pathway for tuning of the magnetic, optical, and electronic response of topological heterostructures. We also observed large changes in the \BS\ phonons due to strain induced by EuS. This confirms a strong elastic coupling between the materials that could be exploited in future devices.

\bibliography{References}

\end{document}